# A Structural Insight into Mechanical Strength of Graphene-like Carbon and Carbon Nitride Networks

Obaidur Rahaman*[,1] Bohayra Mortazavi*[,1], Arezoo Dianat[2], Gianaurelio Cuniberti[2], and  Timon Rabczuk*[,3]

[1]*Institute of Structural Mechanics, Bauhaus-Universität Weimar, Marienstr. 15, D-99423 Weimar, Germany*

[2]*Institute for Materials Science and Max Bergmann Center of Biomaterials, TU Dresden, D-01062 Dresden, Germany*

[3]*College of Civil Engineering, Department of Geotechnical Engineering, Tongji University*

Corresponding author:

Tel: +49 3643584521    E-Mail:  ramieor@gmail.com (Obaidur Rahaman)

Tel: +4915780378770   E-Mail: bohayra.mortazavi@gmail.com (Bohayra Mortazavi)

Tel: +49 3643 584511   E-Mail:  timon.rabczuk@uni-weimar.de (Timon Rabczuk)

## Abstract

Graphene, one of the strongest materials ever discovered, triggered the exploration of many 2D materials in the last decade. However, the successful synthesis of a stable nanomaterial requires a rudimentary understanding of the relationship between its structure and strength. In the present study, we investigate the mechanical properties of 8 different carbon-based 2D nanomaterials by performing extensive density functional theory calculations. The considered structures were just recently either experimentally synthesized or theoretically predicted. The corresponding stress-strain curves and elastic moduli are reported. They can be useful in training force field parameters for large scale simulations. A comparative analysis of these results revealed a direct relationship between atomic density per area and elastic modulus. Furthermore, for the networks that have an armchair and a zigzag orientation, we observed that they were more stretchable in the zigzag direction than the armchair direction. A critical analysis of the angular distributions and radial distribution functions suggested that it could be due to the higher ability of the networks to suppress the elongations of the bonds in the zigzag direction by deforming the bond angles. The structural interpretations provided in this work not only improve the general understanding of a 2D material's strength but also enables us to rationally design them for higher qualities.

## 1. Introduction

Graphene is a one-atom thick, two-dimensional allotrope of carbon with remarkable properties. It is one of the strongest materials ever discovered[1] and an efficient conductor of heat[2] and electricity.[3] Ever since its successful synthesis in 2004,[4] the scientific interest in graphene is rapidly growing due to its current and potential applications in electronics, nano devices, composites, photovoltaics, energy storage etc.[5-14]

Graphene is also the basic structure of many other carbon based allotropes. For instance the stable form of graphite[15] is a stack of alternately shifted graphene sheets, the one dimensional carbon nanotubes[16] can be thought as a graphene sheet rolled into a cylinder and the zero dimensional fullerenes[17] as a graphene sheet rolled into a sphere. In addition to these graphene-like structures, during the last couples of years the scientific community has proposed a myriad of other carbon based 2D structures with interesting electric and mechanical properties, for example, phagraphene,[18] amorphized graphene,[19, 20] biphenylene,[21] graphenylene, [22-24] Haeckelites[25-27] etc. Since the application of graphene in nanoelectronics is limited due to the lack of a band-gap, complex physical or chemical modifications are proposed to overcome this problem.[28-30] Also graphene-like carbon nitride sheets which consist exclusively of covalently-linked $sp^2$-hybridized carbon and nitrogen atoms are experimentally synthesized and have inherent band-gap well-suited for electronic applications.[31-33].

Since the desirable properties of these materials stem from their unique structures, an in depth understanding of their structure-function relationship is essential for improving their qualities as well as performances. Specially, understanding a material's response to mechanical strain at an atomic level is important in designing stable and strong materials. Measuring such properties in carbon nanostructures under experimental conditions can be very challenging. Thus, state of the art, *ab initio* computational methods can be useful in fulfilling this need, especially for novel materials with interesting properties that are not experimentally synthesized yet. Thus, a number of *ab initio* and force field studies were devoted in characterizing the mechanical properties of graphene and other related materials.[34-41]

However, there is a scarcity of computational studies that compare the mechanical properties of different carbon based 2D nanomaterials. Comparing the properties of such related materials can be very useful in putting them into perspective as well as obtaining valuable insight into the structural origin of strength. The knowledge obtained in such studies can be exploited for engineering stronger and more effective materials. They can also be used as benchmarks for parameterizing force fields for performing simulations at a larger scale.

In this work, in order to provide a unique but also a general vision, we selected 8 different carbon and carbon nitride based 2D materials to investigate their responses under uniaxial strain. The structures were selected because of their interesting and desirable properties as well as novelty or successful experimental synthesis. For example, triazine-based graphitic carbon nitride sheets were experimentally fabricated in 2014,[32] nitrogenated holey graphene was synthesized using wet-chemical reaction in 2015[33] and amorphous graphene was experimentally fabricated in 2015.[20] Thus, the mechanical properties of the materials explored in this work are of high relevance and importance.

The mechanical strengths in the elastic region were calculated and compared. As a result of this investigation, a direct relationship between the atomic density of a network and its mechanical strength was established. A comparison of the inelastic regions and breaking points revealed a geometric interpretation of the unequal mechanical responses of materials strained along armchair and zigzag directions.

The knowledge obtained in this work can be useful in understanding the microscopic origin of mechanical strength in a 2D nanomaterial as well as rational design of novel materials with desirable properties.

## 2. Computational method

We used the *Vienna ab Initio Simulation Package[42, 43]* to carry out the *ab initio* calculations within the framework of density functional theory (DFT). The interaction between the frozen core and valence electron was calculated using Perdew, Burke, and Ernzerhof (PBE) functional in conjugation with Projector Augmented Wave (PAW) method.[44] Generalized Gradient Approximation (GGA)[45] was used to treat the exchange-correlation energy. In order to reduce the interaction with its periodic image, a vacuum of length 20Å in the perpendicular direction to the 2D network was introduced. The energy minimizations were carried out using conjugate gradient method with a total energy threshold of $10^{-5}$eV and atomic force components of 0.02 eV/Å. For sampling the Brillouin zone we used an energy cutoff of 450 eV and a k-point mesh size of $7 \times 7 \times 1$ in the Monkhorst-Pack scheme.[46] A higher k-point sampling of $11 \times 11 \times 1$ was used for the calculation of electronic band-gaps. Orthorhombic cells were used for the calculations of all the structures. We used supercells instead of unit cells in our calculations. Since the dynamic effect like temperature is not considered here one could also use unit cells for this calculation. However, structures with some irregularities like defects require the consideration of a large unit cell or supercell. Since we included amorphous graphene with defects in our set of structures it was necessary to use supercells. Thus, for the sake of maintaining uniformity we also used supercells for the other structures.

## 3. Results and Discussions

The supercell structures considered in this work are shown in Figure 1. The simulation box is shown as an outline for each structure. An isotropic response under uniaxial strain is expected from amorphous graphene when significantly large lengths are considered in both X and Y directions. However, this isotropy is difficult to reproduce with a finite supercell size due to its local anisotropy. Thus, four different supercells of similar size are considered for the calculations. Only one of them is shown in Figure 1. The defect concentration in amorphous graphene is defined as the ratio of non-hexagonal rings in the amorphous structure with respect to the total number of hexagons present in the pristine sheet. A 68% defect is considered for all the four cases. The X and Y axes of the simulation box are shown in the figure for amorphous graphene, pentagraphene and phagraphene. For the rest of them an armchair and a zigzag direction were identified and indicated in the figures.

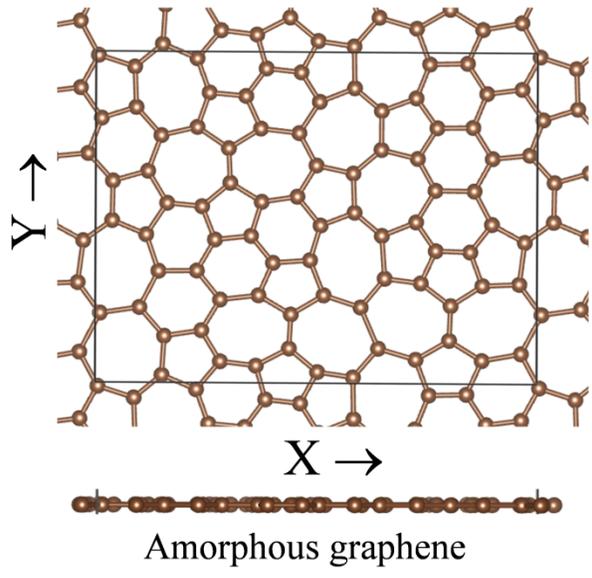

Amorphous graphene

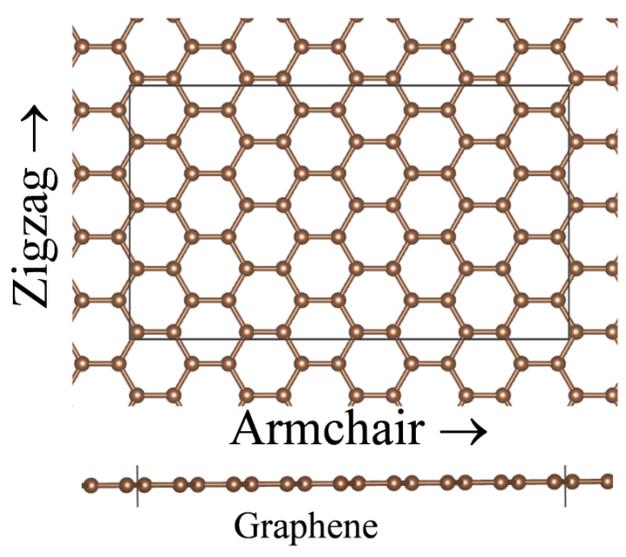

Graphene

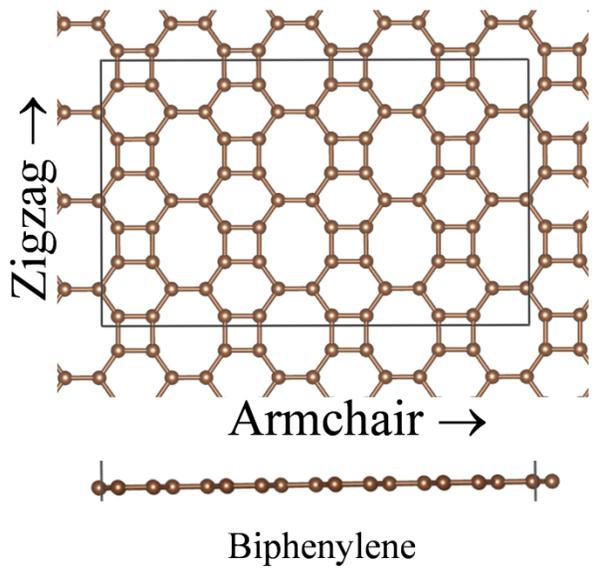

Biphenylene

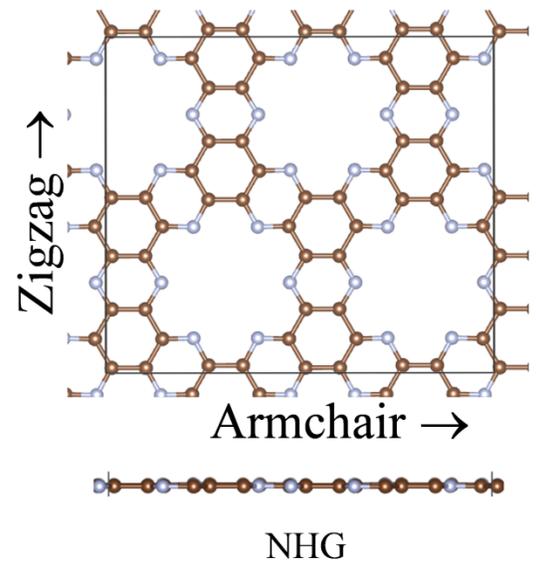

NHG

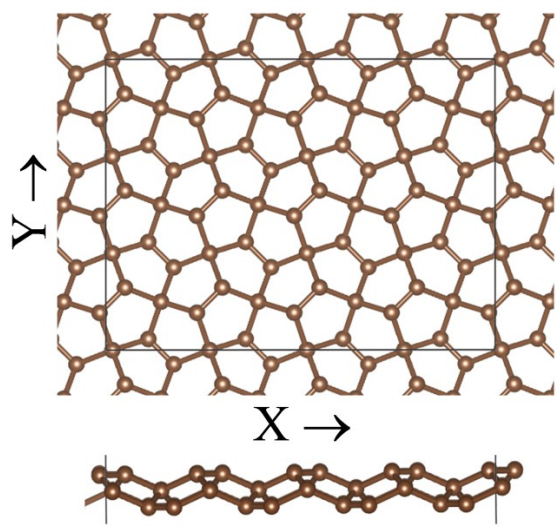

Y →

X →

Pentagraphene

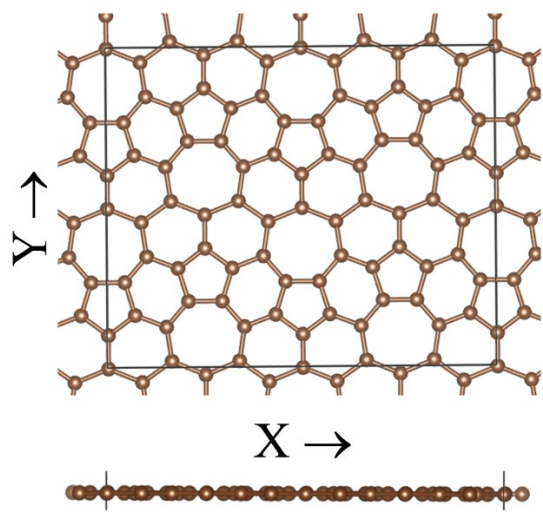

Y →

X →

Phagraphene

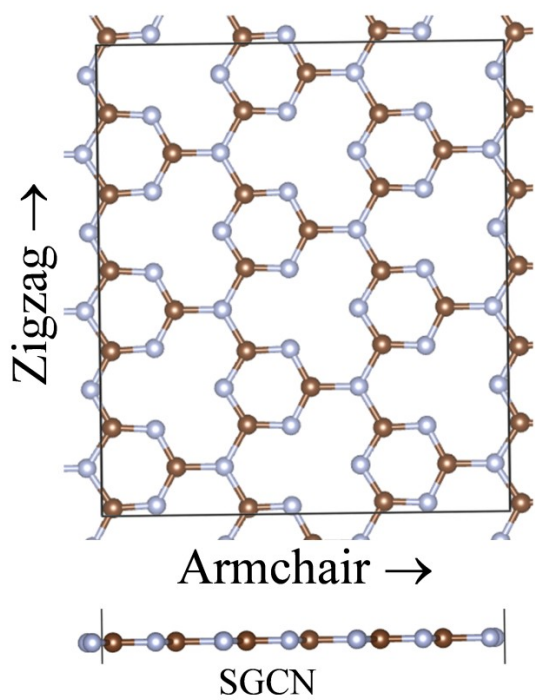

Zigzag →

Armchair →

SGCN

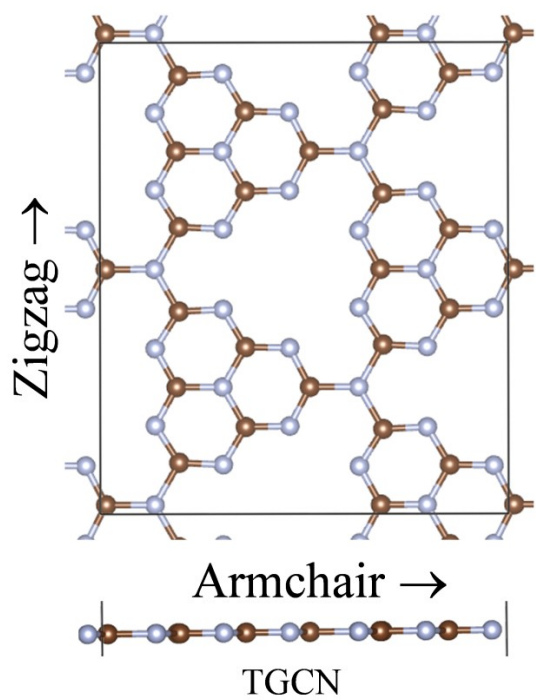

Zigzag →

Armchair →

TGCN

● Carbon    ● Nitrogen

Figure 1: Supercell structures of the 2D carbon and carbon nitride networks.

**3.1 Comparison of the stress-strain curves**

Figure 2 shows the stress-strain curves of the 8 networks subjected to uniaxial loading. The insets show the linear regions and the best-fit lines that were used to calculate the elastic modulus. The maximum stresses attained by the networks in both directions are reported in Table 1. The maximum stress value reported for amorphous graphene was obtained by taking an average of the four sample cases. Graphene attained the highest stress 38.25 GPa.nm in the zigzag direction and 34.25 GPa.nm in the armchair direction. This was followed by phagraphene, pentagraphene, biphenylene, amorphous graphene, nitrogenated holey graphene (NHG),[47] s-triazine-based graphitic carbon nitride (SGCN)[32] and the lowest values of 9.95 GPa.nm in the zigzag direction and 7.34 GPa.nm in the armchair direction for tri-triazine-based graphitic carbon nitride (TGCN).[48] The NHG, SGCN and TGCN structures with significant gaps or holes in the structures were expected to have lower strengths. The Poisson's ratios of the networks were also reported. For all of the networks, the estimated Poisson's ratios did not depend on the direction of stretching. The highest Poisson's ratio of 0.362 was found for biphenylene whereas the lowest Poisson's ratio was -0.85 for pentagraphene. The negative Poisson's ratio of pentagraphene was also observed by the authors who proposed this structure.[49]

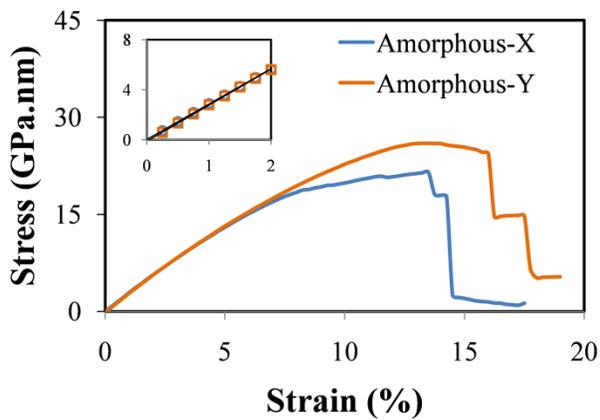

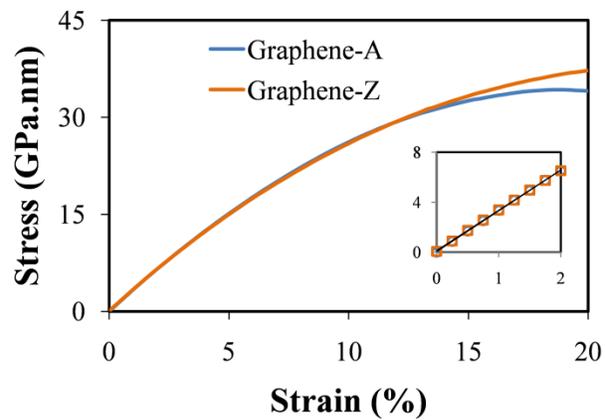

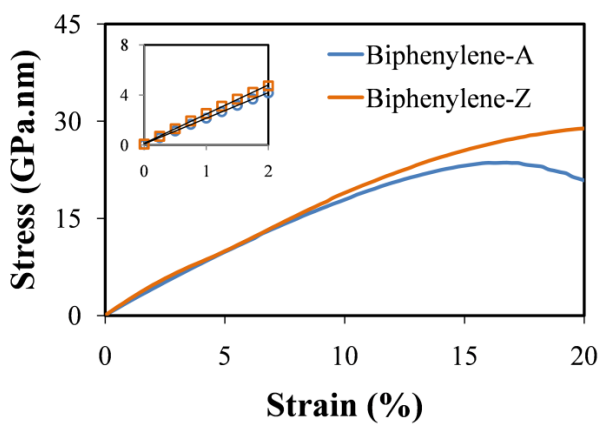

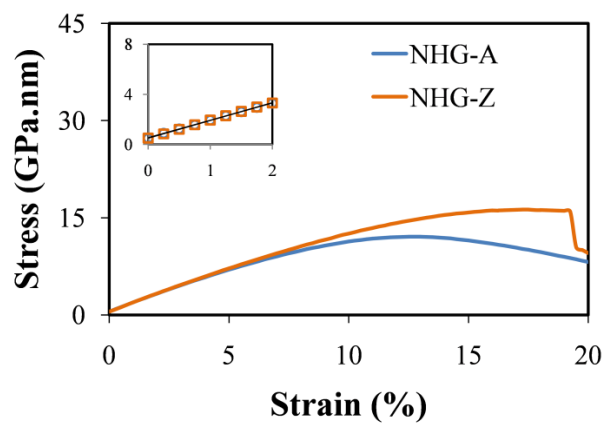

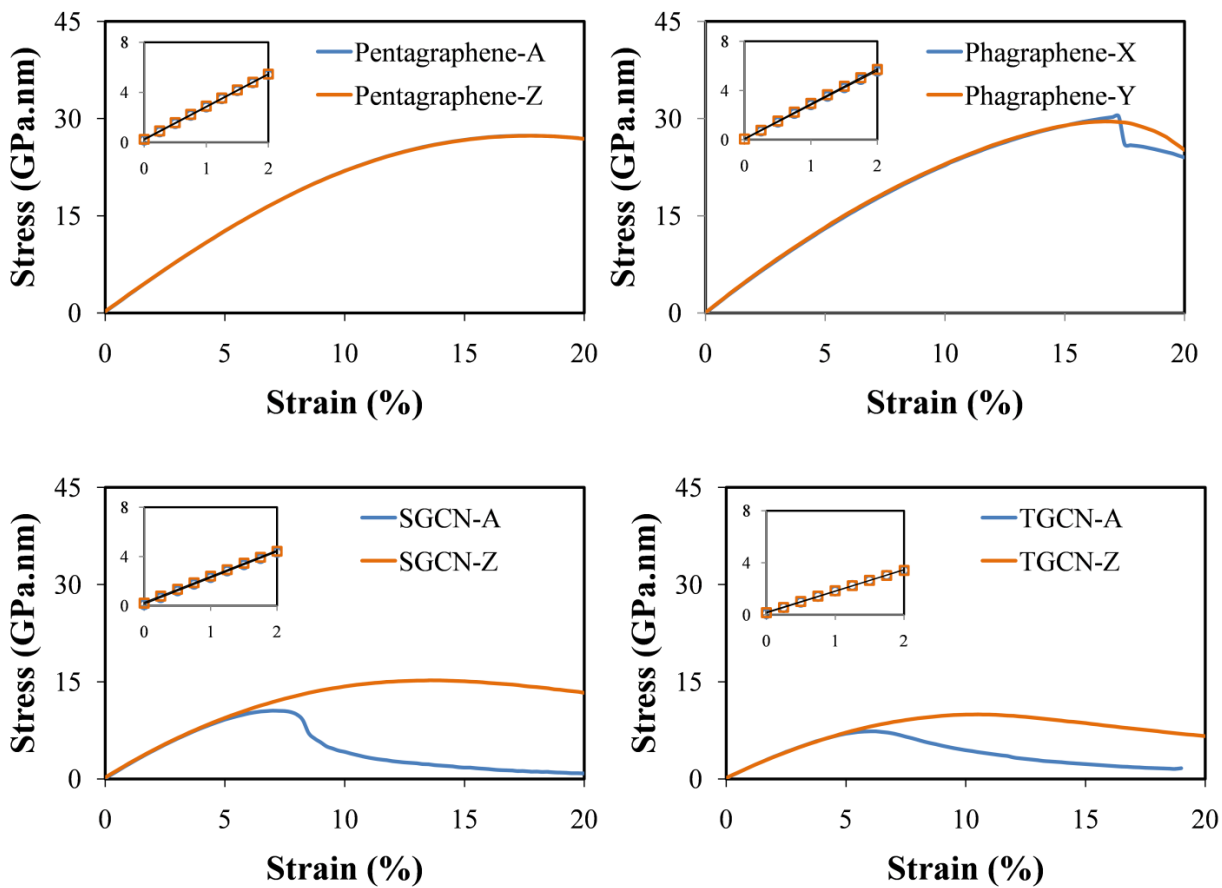

Figure 2: Stress-strain curves of the 8 networks. The insets show the linear regions with the fitted straight lines. The slopes of these lines were used to estimate the elastic modulus.

**Table 1**

| Type | Nbr of atoms | Supercell Size (Å × Å) | Poisson's Ratio | Density Atom/Å2 | $\sigma_{max}$ (Armchair/X) GPa.nm | $\sigma_{max}$ (Zigzag/Y) GPa.nm | E GPa.nm (GPa) |
|---|---|---|---|---|---|---|---|
| Amorphous | 80 | 17.06 × 12.79 | 0.172 ± 0.006 | 0.37 | 20.23 ± 3.96 | 20.23 ± 3.96 | 276 ± 11.1 (809 ± 64 |
| Graphene | 64 | 17.09 × 9.87 | 0.161 ± 0.004 | 0.38 | 34.25 | 38.25 | 325 ± 2.0 (969 ± 6) |
| Biphenylene | 72 | 18.07 × 11.26 | 0.362 ± 0.011 | 0.35 | 23.60 | 29.21 | 220 ± 20.3 (656 ± 61 |
| NHG | 72 | 16.68 × 14.44 | 0.267 ± 0.008 | 0.30 | 12.08 | 16.26 | 140 ± 1.4 (419 ± 5) |
| Pentagraphene | 72 | 14.53 × 10.89 | -0.085 ± 0.002 | – | 27.36 | 27.34 | 262 ± 0.7 (781 ± 2) |
| Phagraphene | 80 | 16.16 × 13.29 | 0.242 ± 0.009 | 0.37 | 29.56 | 29.55 | 281 ± 2.8 (839 ± 8) |
| SGCN | 63 | 12.45 × 14.36 | 0.161 ± 0.008 | 0.35 | 10.53 | 15.22 | 212 ± 0.7 (633 ± 3) |
| TGCN | 56 | 12.34 × 14.25 | 0.233 ± 0.012 | 0.32 | 7.34 | 9.95 | 165 ± 2.1 (492 ± 8) |

The stress-strain curves evolved differently as the networks were strained along a particular direction (Figure 2). Pentagraphene, which is isotropic, produced almost identical stress in both X and Y directions. This was also the case with phagraphene. It is interesting to note that the zigzag direction always produced higher stress than the armchair direction for the networks that were identified with these two types of atomic arrangements. Also, the breaking point was reached at a later stage when they were strained along the zigzag direction, as compared to when they were strained along the armchair direction. In other words, a network containing armchair and zigzag directions is more stretchable in the zigzag direction whereas it breaks earlier (at a lower strain) when stretched along the armchair direction. We have investigated the reason for this trend in some details later.

We note that the mechanical strengths of amorphous graphene,[19] SGCN and TGCN[50] films predicted by classical molecular dynamics (MD) simulations agree well with the results obtained by our DFT calculations. MD simulations, in agreement with our DFT modeling, also predict slightly higher elastic modulus along the zigzag direction in comparison to the armchair direction for pristine graphene.[51]

## 3.2 Evaluation and comparison of elastic modulus

The elastic modulus is a measure of a materials intrinsic strength. Although anisotropy was observed in most of the 2D networks, they almost vanished at a very low strain, with a couple of exceptions. From 0% until 2% strain the stress values looked very similar irrespective of the direction of stretching, except biphenyelene. Thus, except this structure, the network can be considered isotropic at this low level of deformation. The isotropy of graphene under small deformation was predicted by another atomistic simulation study.[52] A close look at the stress-strain curves at this region revealed a linear relationship between them, confirming elastic deformation. The elastic constant E was calculated for the 8 networks in this region (Table 1). The elastic modulus of graphene was estimated to be 325 ± 2.0 GPa.nm or 969 ± 6 GPa. This is the highest among all of them and matches well with previous DFT results.[51] Phagraphene and amorphous graphene also demonstrated significant strength against uniaxial strain. The lowest value of the elastic modulus, 140 ± 1.4 GPa.nm was obtained for the holey structure of NHG. The elastic modulus of pentagraphene was found to be 262 ± 0.7 GPa.nm which matches closely with the value of 263.8 GPa.nm as estimated by another DFT study by Li et al.[53] The errors were estimated by calculating the standard deviations between the elastic moduli obtained along the two directions. The elastic modulus of amorphous graphene was $276 \pm 11.1$ GPa.nm which was calculated by taking the average results of four sample cases, both along the X and Y directions.

### 3.3 The strength is determined by atomic density

Understanding the source of a material's intrinsic strength at a microscopic level is essential in designing future materials of superior strengths and perhaps unique properties. The strength of a material is eventually related to the atomic interactions and chemical bond formations and breakings. The strenght of a 2D network depends both on the number and types of its constituting elements as well as their geometric arrangements. To simplify matters, we focussed only on the reltionship between strength and the number of atoms in the network per unit area.

The cases considered in this work suggest that the elastic modulus increases with higher atomic density (Figure 3). Interestingly this is true irrespective of the geometrical arrangements and atom types of the network,. Although the relationship is not a linear one, the trend is unmistakable. A simple explanation of this could be that the denser a 2D network is the higher is the total number of bonds between the atoms and thus more difficult to be broken by the strain. Following this observation, it can perhaps be safely inferred that the strength of a 2D network is predominantly determined by its atomic density per area and only fine tuned by the types and geometrical arrangements of the constituting atoms. We note that this inference is valid when a number of networks with a broad range of densities, like this work, are compared. For a narrow range of densities, the atom types and geometry can play the major role. We also note that pentagraphene was excluded from this analysis since its structure is non-planar and thus the calculation of its area is not straightforward.

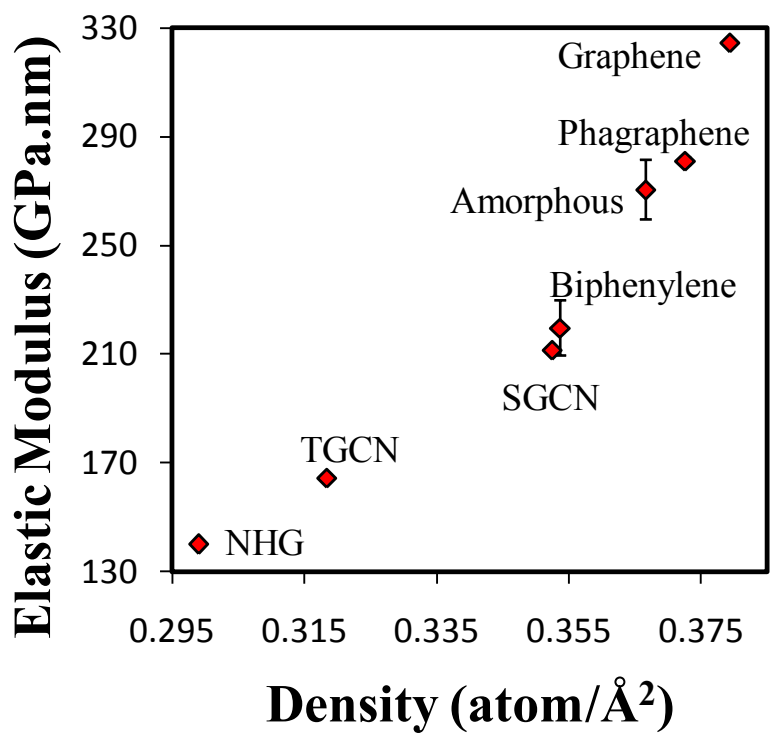

Figure 3: Elastic modulus vs atomic density

## 3.4 Why are the networks more stretchable along the zigzag direction than the armchair direction?

Each of the five networks that have an armchair direction and zigzag direction along its surface (namely graphene, biphenylene, NHG, SGCN and TGCN) demonstrates more stretchability along the zigzag direction. Although, the structural details of an individual network are accountable for its strength in a particular direction, the observed trend might not be entirely a coincidence.

In order to shade light on the microscopic explanation of this phenomenon, we investigated the case of graphene in details. As the 2D carbon network was subjected to uniaxial strain, its response to the strain was manifested by changes in the C-C-C angles as well as the C-C bond lengths. The snapshots of graphene at 0%, 5%, 10%, 15% and 20% uniaxial strain, both in the armchair direction and zigzag direction, were selected. The C-C-C angular distributions and C-C radial distribution functions were calculated at each level of strain (Figure 4 and Figure 5, respectively).

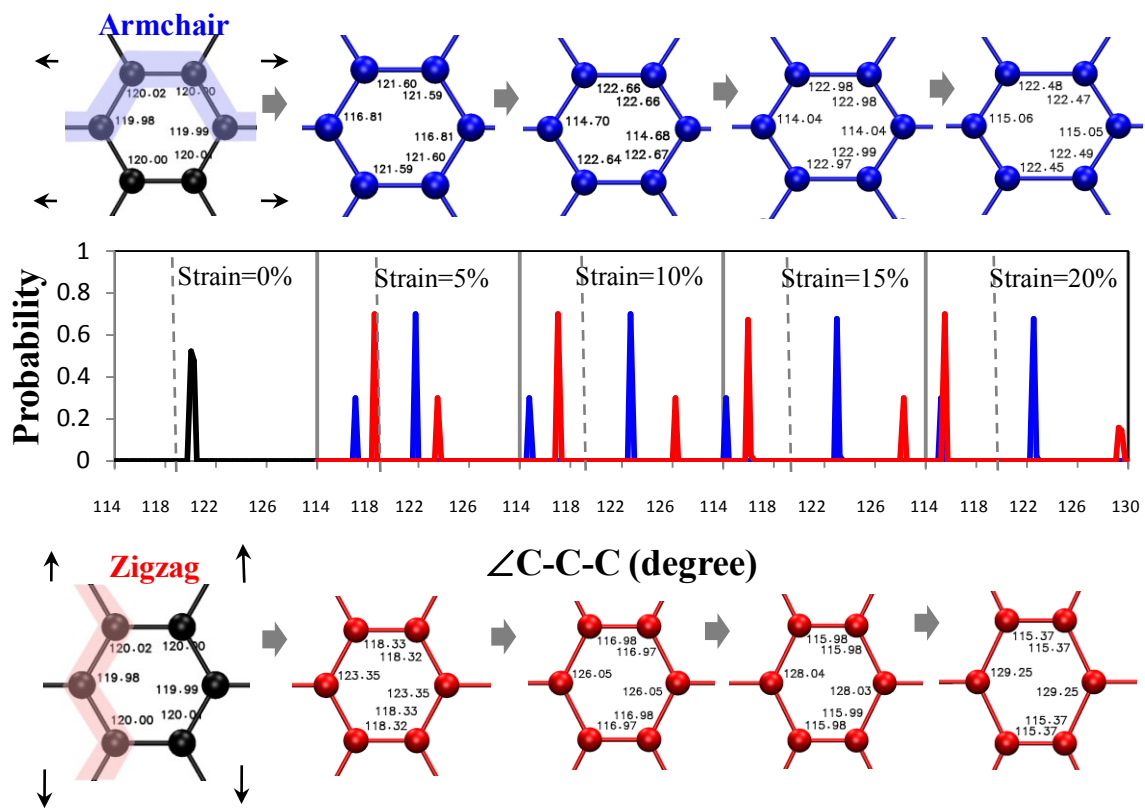

Figure 4: C-C-C angle distributions of graphene strained along the armchair and zigzag directions. The case for the relaxed state is marked by the dotted line.

At the completely relaxed state (strain = 0%), a single peak near 120° was observed in the C-C-C angle distributions of graphene (Figure 4). This corresponds to the most unrestrained geometry of the $SP^2$ hybridization of the C atoms. As soon as the strain is applied, either in the armchair direction or in the zigzag direction, the angles start to deform. Both the structures and the angle distributions for the three cases, namely unrestrained, strained in the armchair direction and strained in the zigzag direction are rendered using black, blue and red colors, respectively. Depending on the atomic positions, a particular C-C-C angle can either increase or decrease in response to the strain, in order to allow the network to stretch. Thus, a splitting of the peak into two is observed for both the strained cases. For each case, the two separated peaks are located on either side of the ideal case of 120° as marked by the black dotted line. Besides these common traits, a notable difference between the armchair and zigzag directions can be observed in the details of their angular distributions. For the armchair case, the peak corresponding to the widening of the angle seems to be smaller than that of the zigzag case. As it can be seen in the sketches, this angle remains around 122° for the whole range of strained cases for the armchair direction. On the other hand, it gradually increases from 123° to about 129° for the zigzag direction. Thus, we can conclude that the angles are more widened while graphene is stretched in the zigzag direction.

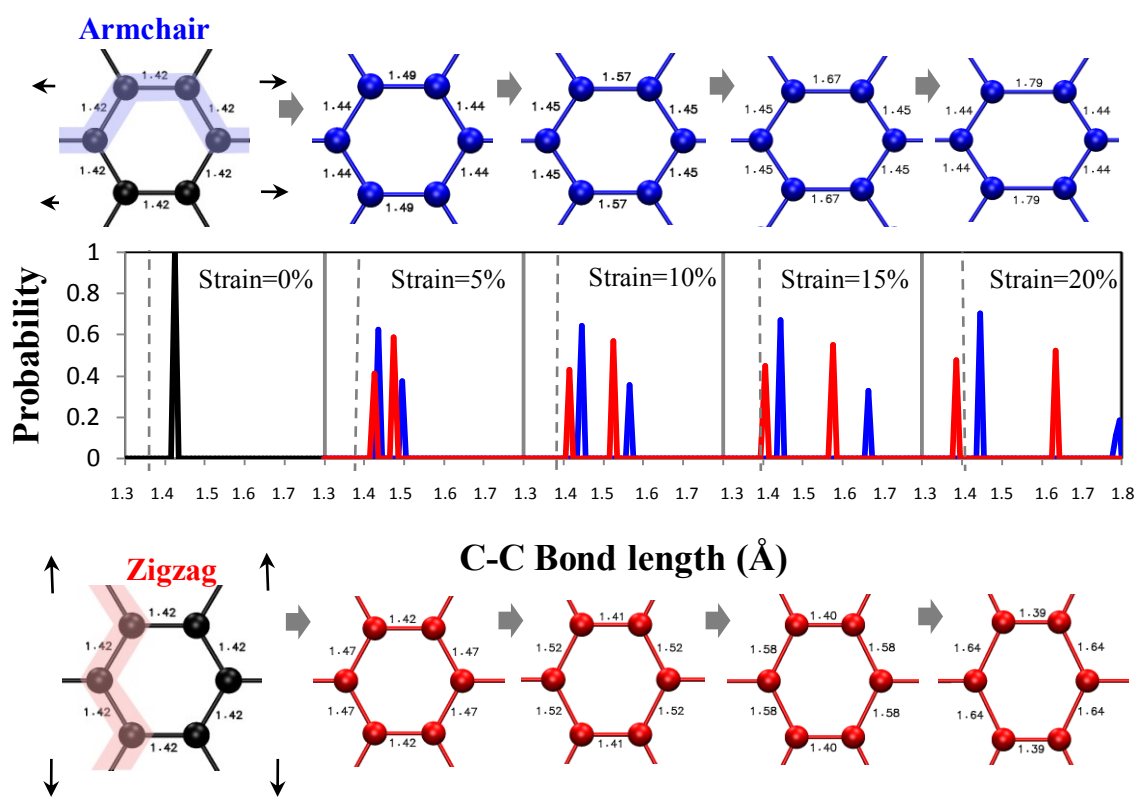

Figure 5: C-C radial distribution functions of graphene strained along the armchair and zigzag directions. The case for the relaxed state is marked by the dotted line.

A single peak at 1.42 was observed in the C-C radial distribution function of graphene at the completely relaxed state (Figure 5). Similar to the observation for angular distribution, the peak split into two in response to the strain applied both in the armchair and zigzag directions. Unlike the angles, the bonds did not shorten significantly. As the stain was gradually increased, the peak corresponding to the shorter bond length slightly increased for the armchair case and slight decreased for the zigzag case. On the other hand the peak corresponding to the longer bond length stretched significantly (from 1.49Å to 1.79Å) for the armchair case but only moderately (from 1.47Å to 1.64Å) for the zigzag case. Thus, we can conclude that the C-C bonds are more stretched when graphene is strained along the armchair direction as compared to the zigzag direction. We note that the stretching of the C-C bonds will eventually tear the material apart. So at strain = 20%, graphene is closer to the breaking point when strained in the armchair direction than the zigzag direction.

Putting the evolutions of C-C-C angular distributions and C-C radial distribution functions together it can be concluded that the higher stretchability of graphene in the zigzag direction can be explained by its ability to slow down the bond stretching by significant widening of the C-C-C angles.

A similar trend was observed for the case of biphenylene (Figure S1 and S2). Due to the presence of different types of rings in its structure, it has four peaks in the C-C-C angular distribution function in the most relaxed state. As the network was stretched in the armchair or the zigzag direction, the first peak at 90° remained in the same position. The distance between the second and the third peak gradually increased when stretched along the armchair direction. On the contrary, it decreased when stretched along the zigzag direction, eventually merging into one peak. The fourth peak corresponding to the most obtuse angle at 145° gradually decreased when stretched along the armchair direction whereas it increased when stretched along the zigzag direction. There are two peaks in the C-C radial distribution function of biphenylene in the relaxed state. These two peaks break into three and eventually four as the network is stretched, both along the armchair and zigzag directions. At a strain of 15%, the fourth peak corresponding to the most stretched C-C bonds remained at 1.74Å for both the cases. However, the third peak corresponding to the second most stretched C-C bonds is at 1.59Å for the armchair case while it is at 1.5Å for the zigzag case. Thus, the armchair case is again closer to the breaking point than the zigzag case.

### 3.5 Strain induced modifications of band-gaps

Uniaxial strain can alter the electronic band-gaps of materials, as observed experimentally[54] or predicted by theoretical studies.[55] The opening of a very small band-gap under uniaxial strain was predicted for graphene.[55] We observed a similar phenomenon in our study. Since it has been discussed earlier in details, we focused our attention on the effect of uniaxial strain on other materials that have band gaps, namely NHG, pentagraphene, SGC and TGCN (Figure 6).

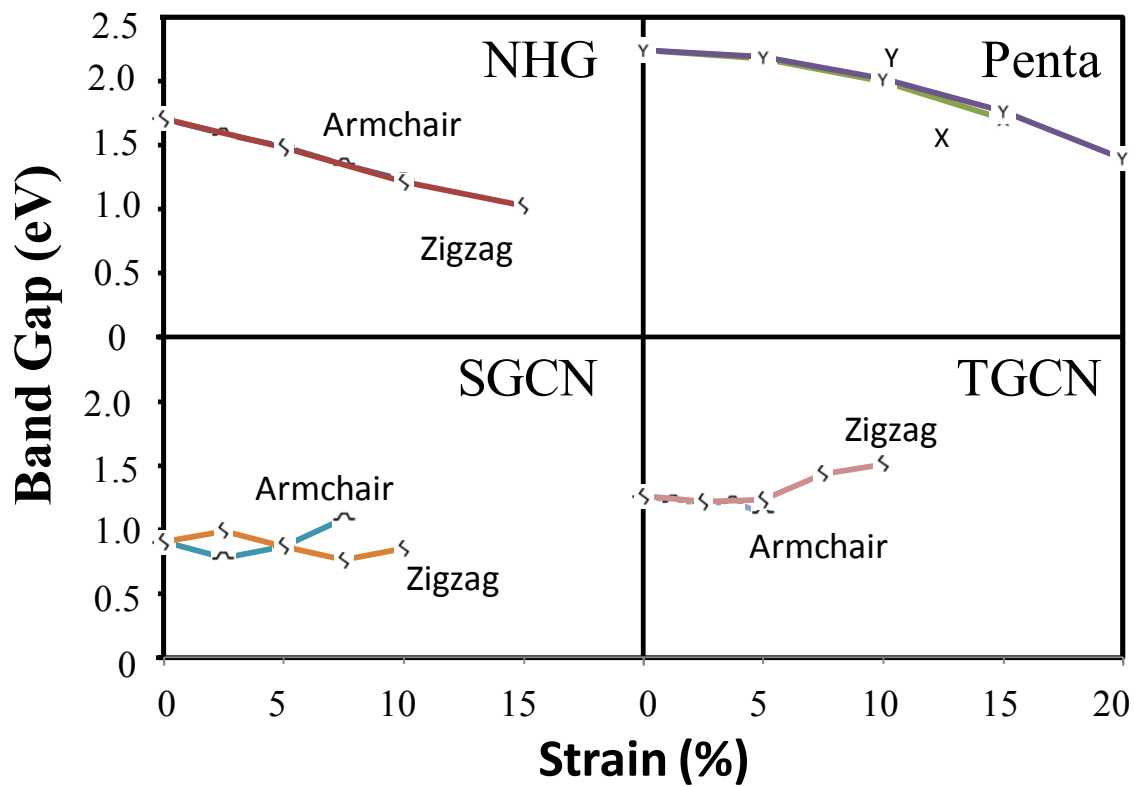

Figure 6: Evolution of band-gaps with uniaxial strain

The band gap gradually reduced with increasing uniaxial strain for both pentagraphene and NHG. For these two cases, the band-gaps were almost identical irrespective of the direction of the strain, armchair or zigzag. For the cases of SGCN and TGCN, slight variations of the band-gaps were observed. For SGCN, the band-gap changed differently depending on the direction of the strain.

## 4. Conclusions

The rapid increase in the number of proposed 2D carbon based networks necessitates not only accurate predictions of their strengths but also a fundamental understanding of the role of structure in determining their strengths. In this work, we selected a wide variety of carbon-based 2D networks and used DFT to measure their mechanical response under uniaxial strain. The stress-strain curves and elastic moduli of the selected networks were reported. They can be useful for comparing the mechanical properties of these materials as well as for training force fields for large scale simulations.

In addition to reporting the mechanical properties of the networks, we established a direct relationship between atomic density per area and elastic modulus. According to this finding, a carbon-based 2D material is likely to have a high elastic modulus if its atomic density per area is high, irrespective of the types of its constituent atoms. This insight can be useful in designing materials with desirable strengths.

For the 2D networks considered in this work the zigzag direction was observed to be more stretchable than the armchair direction. For the case of graphene and biphenylene, a rigorous analysis of the $\angle$C-C-C angular distribution function and the C-C radial distribution function revealed an explanation for this trend. As compared to the armchair direction, the higher stretchability in the zigzag direction is attributed to its higher ability to restrain the elongation of the C-C bonds by deforming the $\angle$C-C-C angles as strain is applied uniaxially.

Furthermore, we reported the modifications of the band-gaps due to uniaxial strain for the 2D networks that have band-gaps. This can be useful in determining the changes in the electronic states of these materials under uniaxial strain.

The results obtained in this work along with the structural insights can be useful in understanding the mechanical properties of 2D networks at a fundamentals level. This understanding can in turn facilitate the design of novel materials with superior strengths.

**Acknowledgements**: The authors OR, BM and TR gratefully acknowledge the financial support of the European Research Council (Grant number 615132).